\documentclass{article}
\usepackage{spconf,amsmath}

\usepackage{multirow}
\usepackage{graphicx}
\usepackage{amsmath}
\usepackage[psamsfonts]{amssymb}
\usepackage{amsxtra}
\usepackage{threeparttable}
\usepackage{amsfonts}
\usepackage{mathrsfs}
\usepackage{arydshln}
\usepackage{cite}
\usepackage{enumitem}
\usepackage{url}

\makeatletter
\let\MYcaption\@makecaption
\makeatother
\usepackage{subcaption}
\makeatletter
\let\@makecaption\MYcaption
\makeatother
\newcommand{\myset}[1]{\mathrm{#1}}

\newcounter{num}
\newcommand{\rnum}[1]{\setcounter{num}{#1} \roman{num}}

\title{Deep Inverse Tone Mapping Using LDR Based Learning
  for Estimating HDR Images with Absolute Luminance}
\name{Yuma Kinoshita and Hitoshi Kiya}
\address{Tokyo Metropolitan University, Tokyo, Japan}
\begin{document}\sloppy
\setlength{\tabcolsep}{1.0pt}
\ninept
\maketitle
\begin{abstract}
  In this paper, a novel inverse tone mapping method using a 
  convolutional neural network (CNN) with LDR based learning
  is proposed.
  In conventional inverse tone mapping with CNNs,
  generated HDR images cannot have absolute luminance, although relative luminance can.
  Moreover, loss functions suitable for learning HDR images are problematic,
  so it is difficult to train CNNs by directly using HDR images.
  In contrast, the proposed method enables us not only to estimate absolute luminance,
  but also to train a CNN by using LDR images.
  The CNN used in the proposed method learns a transformation from various input LDR images to
  LDR images mapped by Reinhard's global operator.
  Experimental results show that HDR images generated by the proposed method
  have higher-quality than HDR ones generated by conventional inverse tone mapping methods,
  in terms of HDR-VDP-2.2 and PU encoding + MS-SSIM.
\end{abstract}
\renewcommand{\thefootnote}{\fnsymbol{footnote}}
\footnote[0]{This work was supported by JSPS KAKENHI Grant Number JP18J20326.}
\renewcommand{\thefootnote}{\arabic{footnote}}
\section{Introduction}
  The low dynamic range (LDR) of modern digital cameras is a major factor
  preventing cameras from capturing images as well as human vision.
  This is due to the limited dynamic range that imaging sensors have.
  For this reason, the interest of high dynamic range (HDR) imaging
  has been increasing.
  
  To generate an HDR image from a single LDR image,
  various research works on inverse tone mapping have so far been reported
  \cite{banterle2006inverse, rempel2007ldr2hdr,
  kuo2012content, huo2013dodging, wang2015pseudo, kinoshita2017fast, kinoshita2017fast_trans,
  endo2017deep, eilertsen2017hdr, marnerides2018expandnet}.
  Traditional way for inverse tone mapping is expanding the dynamic range
  of input LDR images
  by using a fixed function or a specific parameterized function
  \cite{banterle2006inverse, rempel2007ldr2hdr, kuo2012content, huo2013dodging,
  wang2015pseudo, kinoshita2017fast, kinoshita2017fast_trans}.
  However, inverse tone mapping without prior knowledge is generally an ill-posed problem
  because the following two reasons: pixel values might be lost by the sensor saturation,
  a non-linear camera response function (CRF) used at the time of photographing is unknown.
  Hence, HDR images produced by these methods have limited quality.
  To obtain high-quality HDR images,
  inverse tone mapping methods based on deep learning
  have recently attracted attention.

  Several convolutional neural network (CNN) based inverse tone mapping methods
  have been proposed in the past
  \cite{eilertsen2017hdr, endo2017deep, marnerides2018expandnet}.
  These methods significantly improved the performance of inverse tone mapping.
  However, they still have some problems.
  Eilertsen et al. \cite{eilertsen2017hdr} aim to reconstruct saturated areas
  in input LDR images via a CNN using a new loss function calculated in the logarithmic domain,
  but their method is applicable only when a CRF used at the time of photographing is given.
  Endo et al. \cite{endo2017deep} have proposed a CNN based method
  that produce a set of differently exposed images from a single LDR image,
  to avoid difficulty of desining loss functions for HDR images.
  In the work by Marnerides et al. \cite{marnerides2018expandnet},
  an HDR image is directly produced by a CNN,
  where all HDR images are normalized into the interval $[0, 1]$ instead of designing
  loss functions for HDR images.
  HDR images produced by Endo's and Marnerides' methods cannot have absolute luminance,
  although relative luminance can.
  Therefore, luminance calibration is necessary for all predicted HDR images.
  
  Thus, in this paper, we propose a novel inverse tone mapping method
  using a CNN with LDR based learning.
  The proposed method enables us not only to estimate absolute luminance,
  but also to train a CNN by using LDR images.
  Instead of learning a map from LDR images to HDR ones,
  the CNN used in the proposed method learns a transformation from various input LDR images to
  LDR images mapped by Reinhard's global operator.
  Our inverse tone mapping is done
  by applying the inverse transform of Reinhard's global operator
  to LDR images produced by CNNs.

  We evaluated the effectiveness of the proposed method
  in terms of the quality of generated HDR images,
  by a number of simulations.
  In the simulations, the proposed method was compared with
  state-of-the-art inverse tone mapping methods.
  Visual comparison results show that the proposed method can produce
  higher-quality HDR images than that of conventional methods.
  Moreover, the proposed method outperforms the conventional methods
  in terms of two objective quality metrics: HDR-VDP-2.2 and PU encoding + MS-SSIM.
  
\section{Preparation}
\begin{figure}[t]
    \centering
    \includegraphics[clip, width=\columnwidth]{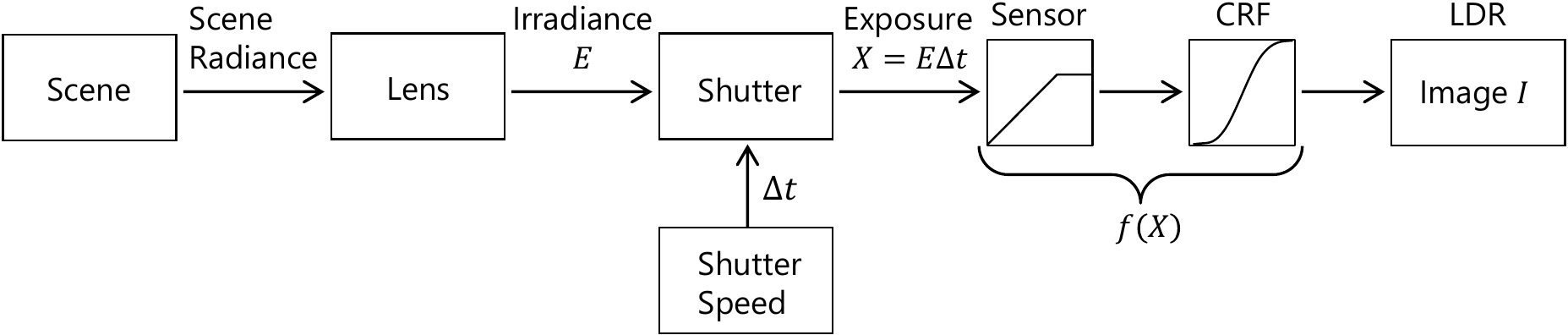}
    \caption{Imaging pipeline of digital camera \label{fig:camera}}
  \end{figure}

  Figure \ref{fig:camera} shows a typical imaging pipeline for a digital camera
  \cite{dufaux2016high}.
  HDR images have pixel values that denote the radiant power density at the sensor,
  i.e., irradiance $E$.
  The goal of inverse tone mapping is restoring irradiance $E$
  from a single LDR image $I$ distorted by the sensor saturation and a non-linear CRF.

\subsection{Tone mapping}
  Tone mapping is an operation that generates an LDR image from an HDR image.
  Since HDR images correspond to irradiance $E$,
  tone mapping can be interpreted as virtually taking photos.
  
  Reinhard's global operator is one of typical tone mapping methods
  \cite{reinhard2002photographic}.
  Under the use of the operator, each pixel value in LDR image $I$
  is calculated from HDR image $E$ by
  \begin{align}
    I_{i, j} &= \hat{f}(X_{i, j}), \label{eq:map} \\
    \hat{f}(X_{i, j}) &= \frac{X_{i, j}}{1+X_{i, j}}, \label{eq:tmo}
  \end{align}
  where $(i, j)$ denotes a pixel and
  $X_{i, j}$ is given by using two parameters $a$ and $G(E)$ as
  \begin{equation}
    X_{i, j} = \frac{a}{G(E)} E_{i, j}. \label{eq:scale}
  \end{equation}
  In Reinhard's global operator, two parameters $a$ and $G(E)$ are used.
  $a \in [0, 1]$ determines brightness of an output LDR image $I$
  and $G(E)$ is the geometric mean of $E$, given by
  \begin{equation}
    G(E) = 
      \exp{
        \left(\frac{1}{|\myset{P}|}
          \sum_{(i, j) \in \myset{P}}
          \log{\left(\max{\left( E_{i, j}, \epsilon \right)}\right)}
        \right)
      },
    \label{eq:geoMeanEps}
  \end{equation}
  where $\myset{P}$ is the set of all pixels and
  $\epsilon$ is a small value to avoid singularities at $E_{i, j} = 0$.

\subsection{Inverse transform of Reinhard's global operator}
  Reinhard's global operator is invertible when two conditions are satisfied:
  two parameters $a$ and $G(E)$ are given, and LDR image $I$ is not quantized.
  The inverse transform is given from eqs. (\ref{eq:map}), (\ref{eq:tmo}), and (\ref{eq:scale},
  as follows:
  \begin{align}
    E'_{i, j} &= \frac{G(E)}{a}X'_{i, j}, \label{eq:inv_scale}\\
    X'_{i, j} &= \hat{f}^{-1}(I_{i, j}) = \frac{I_{i, j}}{1 - I_{i, j}}. \label{eq:inv_tmo}
  \end{align}
  Therefore, when LDR images are generated by Reinhard's global operator,
  HDR images can be reconstructed under the conditions,
  so the HDR images can have the same absolute luminance as that of the original HDR ones.
  
  The literature \cite{kinoshita2017fast_trans} showed that
  parameter $G(E)$ can be calculated by using parameter $a$ and LDR image $I$, as
  \begin{equation}
    G(E) =
      \exp{\left(
        \frac{|\myset{P}|}{|\myset{P}_B|}\log{G(X')}
        - \frac{|\overline{\myset{P}_B}|}{|\myset{P}_B|}\log{a}
      \right)},
      \label{eq:geoMeanG}
  \end{equation}
  where $G(X')$ is calculated by substituting $X'_{i, j}$ of eq. (\ref{eq:inv_tmo})
  into eq. (\ref{eq:geoMeanEps}),
  $\myset{P}_B = \{(i, j) \in \myset{P} | I_{i, j} = 0\}$,
  and $\overline{\myset{P}_B} = \{(i, j) \in \myset{P} | I_{i, j} \neq 0\}$.
  Therefore, the inverse transform of Reinhard's global operator is done
  by using only $a$.

  However, ordinary LDR images are not produced by using Reinhard's global operator.
  In this paper, we aim to transform any LDR images into LDR ones
  produced by Reinhard's global operator, by using a CNN.
  This conception will lead to a novel inverse tone mapping method
  which consists of the CNN and the inverse transform of Reinhard's global operator.

\section{Proposed inverse tone mapping}
  The proposed inverse tone mapping operation is described here.

\subsection{Overview}
\begin{figure}[!t]
    \centering
    \includegraphics[width=\columnwidth]{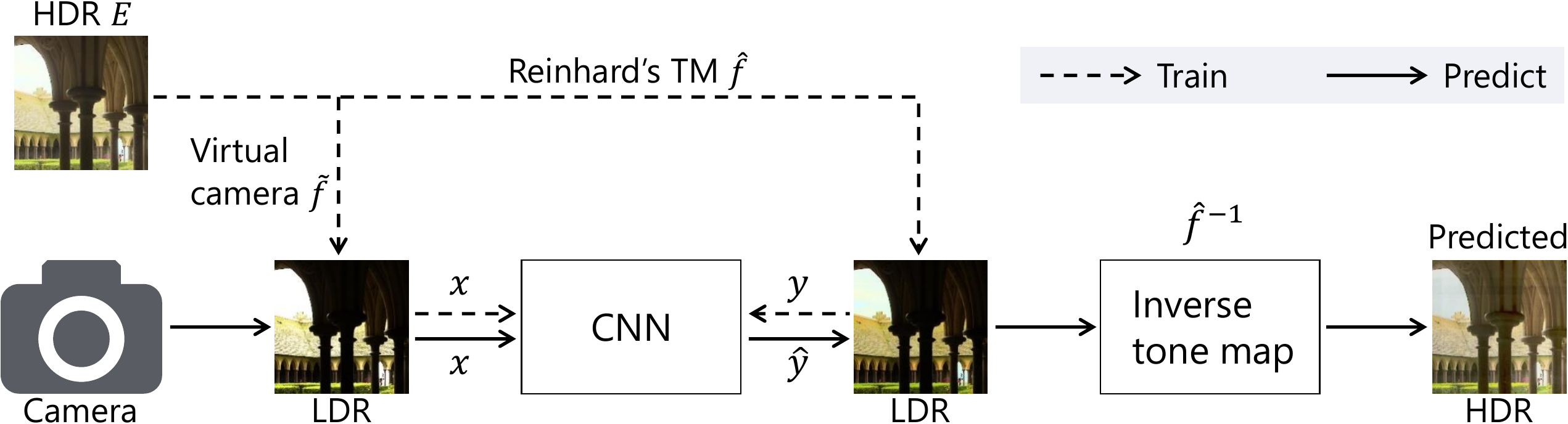}
    \caption{Proposed inverse tone mapping \label{fig:scheme}}
  \end{figure}

  The following is an overview of our training procedure and predicting procedure
  (see Fig. \ref{fig:scheme}).
\subsubsection*{\textbf{Training}}
  \begin{enumerate}[label=\roman*, align=parleft, leftmargin=*, nosep]
    \item Generate input LDR image $x$ from HDR image $E$,
      by using a virtual camera with various CRFs.
      This corresponds to assuming that input LDR images are captured with various cameras.
    \item Generate target LDR image $y$ from HDR image $E$ by using Reinhard's global operator
      with parameter $a = 0.18$, by using eq. (\ref{eq:map}) to eq. (\ref{eq:geoMeanEps}),
      where $a = 0.18$ maps the average luminance of $y$ to the middle gray
      \cite{reinhard2002photographic}.
    \item Train a CNN in order to transform input LDR image $x$ into
      target LDR image $y$.
  \end{enumerate}
  Detailed training conditions will be shown in 3.4.
\subsubsection*{\textbf{Predicting}}
  \begin{enumerate}[label=\roman*, align=parleft, leftmargin=*, nosep]
    \item Let $x$ be an LDR image taken with a digital camera.
    \item Transform $x$ into $\hat{y}$ by using the trained CNN.
      This transformation aims to estimate
      an LDR image generated by Reinhard's global operator.
    \item Generate an HDR image from $\hat{y}$
      by using eq. (\ref{eq:inv_scale}) to eq. (\ref{eq:geoMeanG}),
      where parameter $a = 0.18$ is also utilized in inverse tone mapping.
  \end{enumerate}

  In the proposed method, all target LDR images $y$ are generated with a fixed parameter
  $a = 0.18$.
  Therefore, parameter $a = 0.18$ can be used when the inverse transform of
  Reinhard's global operator is applied to LDR images $\hat{y}$ generated by our CNN.
  As described in 2.2, another parameter $G(E)$ can be calculated
  by using eq. (\ref{eq:geoMeanG}) with $a$ and LDR image $\hat{y}$,
  so HDR images with absolute luminance can be estimated.

\subsection{Network architecture}
\begin{figure}[t]
    \centering
    \includegraphics[clip, width=0.95\hsize]{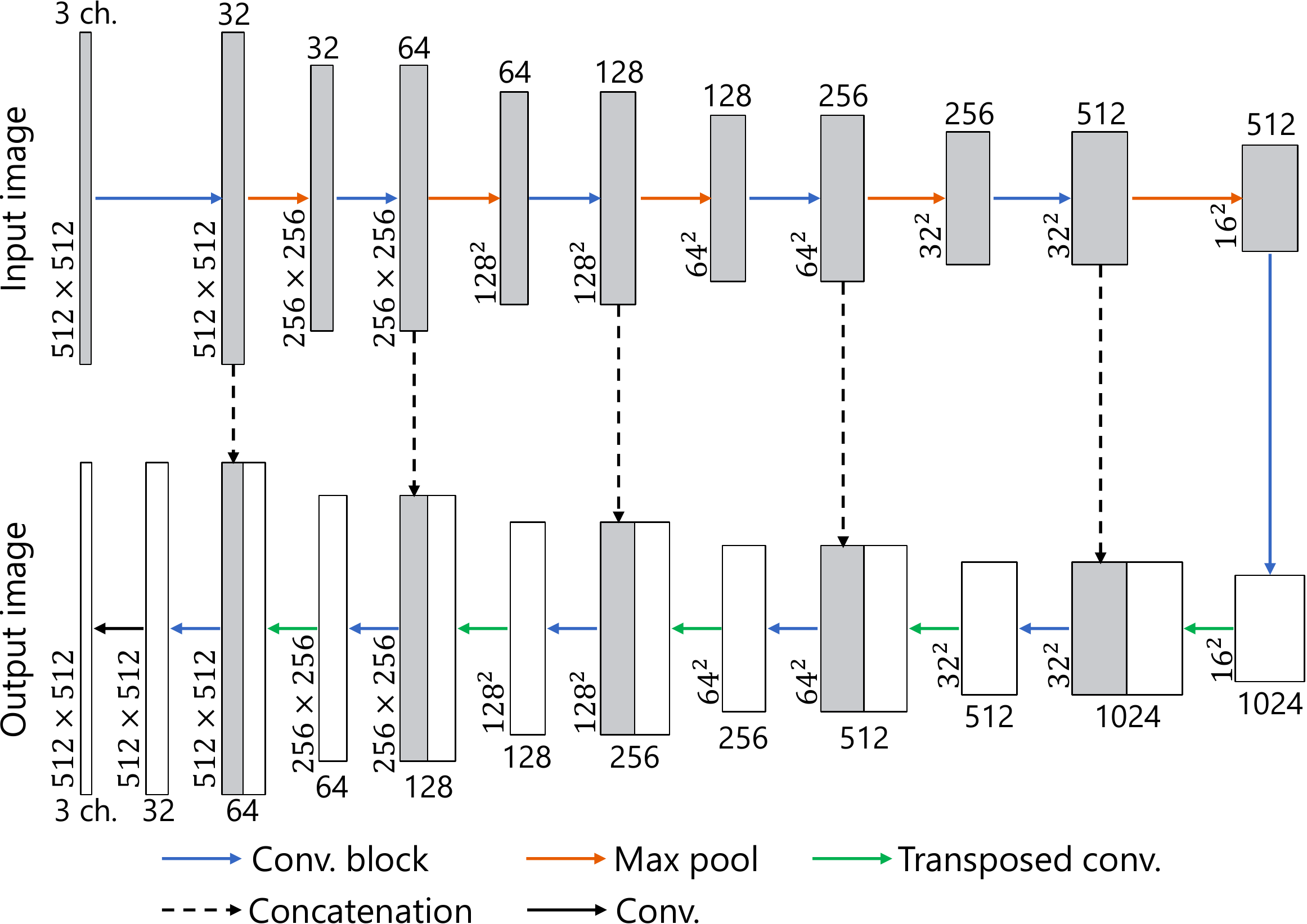}
    \caption{Network architecture.
      Each box denotes multi-channel feature map produced by each layer (or block).
      Number of channels is denoted on top or bottom of each box.
      Resolution of feature map is provided at left edge of box.
      \label{fig:network}}
  \end{figure}

  Figure \ref{fig:network} shows the network architecture of the CNN
  used in the proposed method.
  The CNN is designed on the basis of U-Net\cite{ronneberger2015unet}.
  The input of this CNN is a 24-bit color LDR image
  with a size of $512 \times 512$ pixels.

  Each convolutional block consists of two convolutional layers,
  in which the number of filters $K$ is commonly the same.
  From the first block to the last block,
  the numbers are given as $K = 32$, $64$, $128$, $256$, $512$, $1024$, $512$,
  $256$, $128$, $64$, and $32$,
  where all filters in convolutional blocks have a size of $3 \times 3$.
  Max pooling layers with a kernel size of $2 \times 2$ and
  a stride of $2$ are utilized for image downsampling.
  
  For image upsampling, transposed convolutional layers with a stride of $2$
  and a filter size of $4 \time 4$ are used in the proposed method.
  From the first transposed convolutional layer to the last one,
  the numbers of filters are $K = 512$, $256$, $128$, $64$,
  and $32$, respectively.
  Finally, an output LDR image is produced by
  a convolutional layer which has three filters with a size of $3 \times 3$.

  The rectified linear unit (ReLU) activation function \cite{glorot2011deep}
  is used for all convolutional layers and transposed convolutional ones
  except the final convolutional layer.
  Further, batch normalization \cite{ioffe2015batch} is applied to
  outputs of ReLU functions after convolutional layers.
  The activation function of the final layer is a sigmoid function.

\subsection{Training}
  A lot of LDR images $x$ taken under various conditions
  and corresponding LDR ones $y$ generated by Reinhard's
  global operator are needed for training the CNN in the proposed method.
  However, collecting these images with a sufficient amount is difficult.
  We therefore utilize HDR images $E$ to generate both input images $x$
  and target images $y$ by using a virtual camera \cite{eilertsen2017hdr}
  and Reinhard's global operator, respectively.
  For training, total 978 HDR images was collected from online available databases
  \cite{openexrimage, anyherehdrimage, hdrps, empa, nemoto2015visual, maxplanck}.
  
  A training procedure of our CNN is shown as follows.
  \begin{enumerate}[label=\roman*, align=parleft, leftmargin=*, nosep]
    \item Select eight HDR images from 978 HDR images at random.
    \item Generate total eight couples of input LDR image $x$ and target LDR one $y$
      from each of the eight HDR ones.
      Each couple is generated according to the following steps.
      \begin{enumerate}[label=(\alph*), align=parleft, leftmargin=*, nosep]
        \item Crop an HDR image $E$ to an image patch $\tilde{E}$ with $N \times N$ pixels.
          The size $N$ is given as a product of a uniform random number in range $[0.2, 0.6]$
          and the length of a short side of $E$.
          In addition, the position of the patch in the HDR image $E$
          is also determined at random.
        \item Resize $\tilde{E}$ to $512 \times 512$ pixels.
        \item Flip $\tilde{E}$ upside down with probability 0.5.
        \item Flip $\tilde{E}$ left and right with probability 0.5.
        \item Calculate exposure $X$ from $\tilde{E}$ by
          $X_{i, j} = \Delta t \cdot \tilde{E}$,
          where shutter speed $\Delta t$ is calculated as
          $\Delta t = 0.18 \cdot 2^v / G(\tilde{E})$
          by using a uniform random number $v$ in range $[-4, 4]$.
        \item Generate an input LDR image $x$ from $X$
          by a virtual camera $\tilde{f}$, as
          \begin{equation}
            x = \tilde{f}(X) = \min \left((1+\eta)\frac{X^\gamma}{X^\gamma + \eta}, 1 \right),
          \end{equation}
          where $\eta$ and $\gamma$ are random numbers
          that follow normal distributions with mean $0.6$ and variance $0.1$
          and with mean $0.9$ and variance $0.1$, respectively.
        \item Generate a target LDR image $y$ from $\tilde{E}$
          by Reinhard's global operator (see eq. (\ref{eq:map}) to eq. (\ref{eq:geoMeanEps}))
          with parameter $a = 0.18$.
      \end{enumerate}
    \item Predict eight LDR images $\hat{y}$ from eight input LDR images $x$ by the CNN.
    \item Evaluate errors between predicted images $\hat{y}$ and target images $y$
      by using the mean squared error.
    \item Update filter weights $\omega$ and biases $b$ in the CNN by back-propagation.
  \end{enumerate}
  Note that steps \rnum{2}(f) and \rnum{2}(g) are applied to luminance of $\tilde{E}$,
  and then RGB pixel values of $x$ and $y$ are obtained
  so that ratios of RGB values of LDR images are equal to those of HDR images.

  In our experiments, the CNN was trained with 5000 epochs,
  where the above procedure was repeated 122 times in each epoch.
  In addition, each HDR image had only one chance to be selected,
  in step \rnum{1} in each epoch.
  He's method \cite{he2015delving}
  was used for initialization of the CNN
  In addition, Adam optimizer \cite{kingma2014adam} was utilized for optimization,
  where parameters in Adam were set as $\alpha=0.002, \beta_1=0.5$ and $\beta_2=0.999$.

\section{Simulation}
  We evaluated the effectiveness of the proposed method
  by using two objective quality metrics.

\subsection{Simulation conditions}
  In this experiment, test LDR images was generated
  from seven HDR images which were not used for training,
  according to steps from \rnum{2}(a) to \rnum{2}(f) in 3.3.

  The quality of HDR images $\tilde{E}'$ generated by the proposed method
  is evaluated by two objective quality metrics:
  HDR-VDP-2.2 \cite{narwaria2015hdr},
  and PU encoding \cite{aydin2008extending} with MS-SSIM \cite{wang2003multiscale}
  which utilize an original HDR image $\tilde{E}$ as a reference.
  Literature \cite{hanhart2015benchmarking} have showed that
  these metrics are suitable for evaluating the quality of HDR images.

  The proposed method is compared with three state-of-the-art methods:
  direct inverse tone mapping operator (Direct ITMO) \cite{kinoshita2017fast_trans},
  pseudo-multi-exposure-based tone fusion (PMET) \cite{wang2015pseudo},
  deep reverse tone mapping (DrTMO) \cite{endo2017deep},
  where the third method is based on CNNs,
  but the other methods are not based on machine learning.
\subsection{Results}
\begin{figure*}[!t]
    \centering
    \begin{subfigure}[t]{0.16\hsize}
      \centering
      \includegraphics[width=\columnwidth]{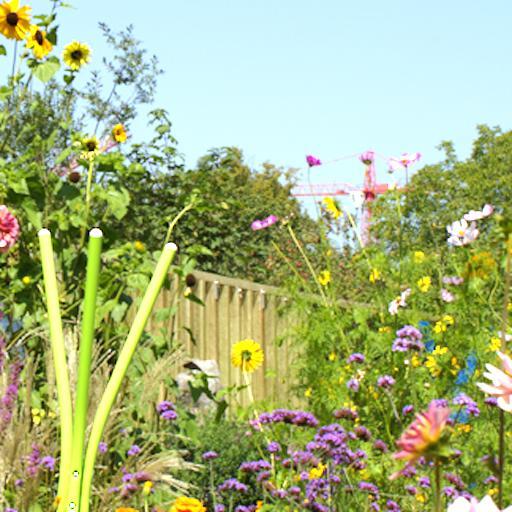}
      \caption{Input $x$ \label{fig:input1}}
    \end{subfigure}
    \begin{subfigure}[t]{0.16\hsize}
      \centering
      \includegraphics[width=\columnwidth]{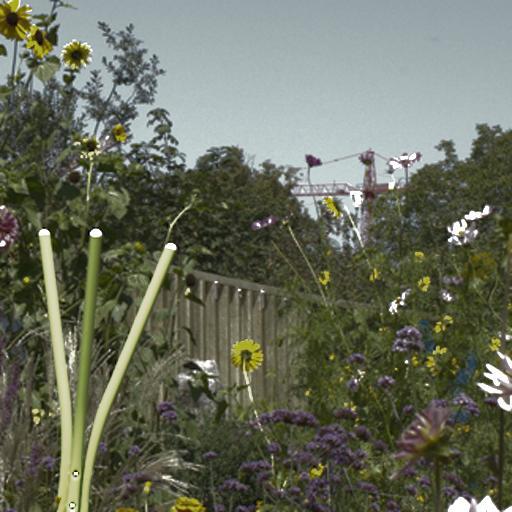}
      \caption{Direct ITMO\cite{kinoshita2017fast_trans} \label{fig:direct1}}
    \end{subfigure}
    \begin{subfigure}[t]{0.16\hsize}
      \centering
      \includegraphics[width=\columnwidth]{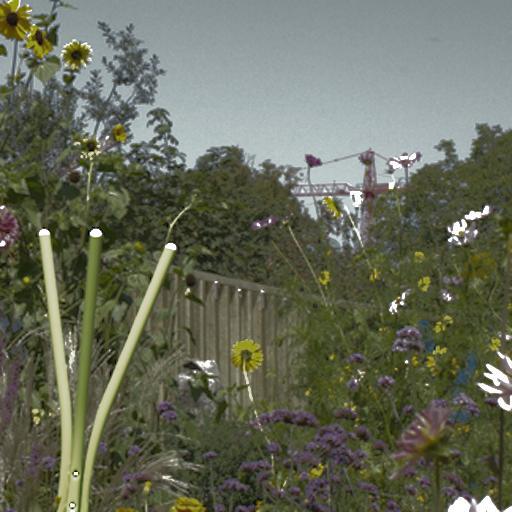}
      \caption{PMET\cite{wang2015pseudo} \label{fig:pmet1}}
    \end{subfigure}
    \begin{subfigure}[t]{0.16\hsize}
      \centering
      \includegraphics[width=\columnwidth]{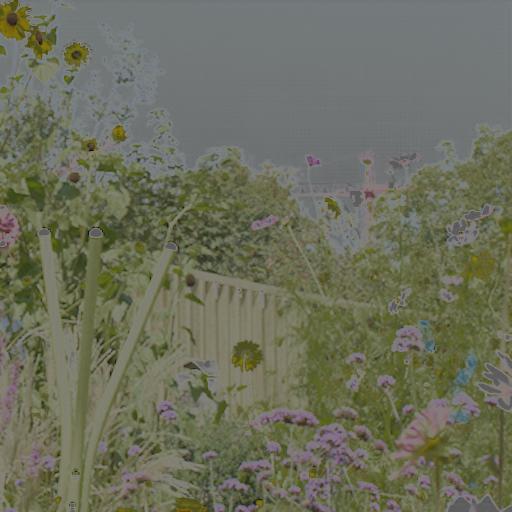}
      \caption{DrTMO\cite{endo2017deep} \label{fig:drtmo1}}
    \end{subfigure}
    \begin{subfigure}[t]{0.16\hsize}
      \centering
      \includegraphics[width=\columnwidth]{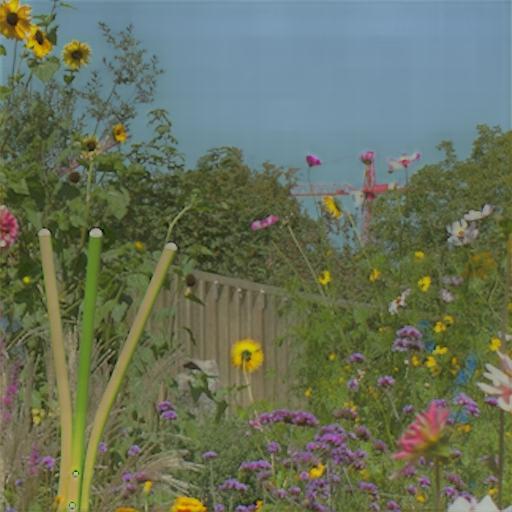}
      \caption{Proposed \label{fig:proposed1}}
    \end{subfigure}
    \begin{subfigure}[t]{0.16\hsize}
      \centering
      \includegraphics[width=\columnwidth]{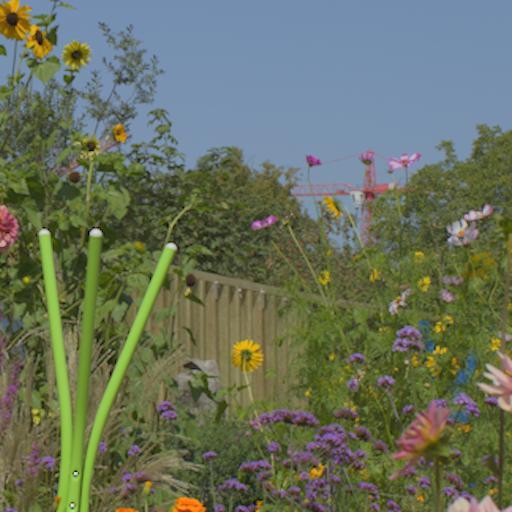}
      \caption{Ground truth $\tilde{E}$ \label{fig:original1}}
    \end{subfigure}
    \caption{Experimental Results (Image 1) \label{fig:result1}}
  \end{figure*}

\begin{figure*}[!t]
    \centering
    \begin{subfigure}[t]{0.16\hsize}
      \centering
      \includegraphics[width=\columnwidth]{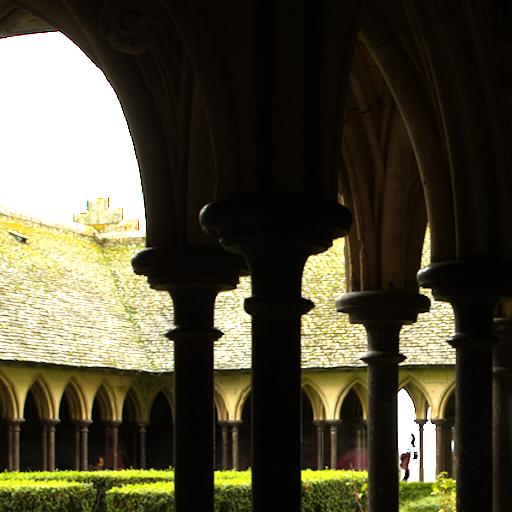}
      \caption{Input $x$ \label{fig:input5}}
    \end{subfigure}
    \begin{subfigure}[t]{0.16\hsize}
      \centering
      \includegraphics[width=\columnwidth]{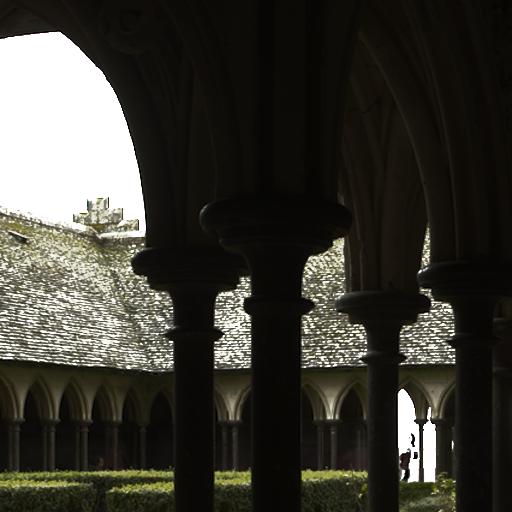}
      \caption{Direct ITMO\cite{kinoshita2017fast_trans} \label{fig:direct5}}
    \end{subfigure}
    \begin{subfigure}[t]{0.16\hsize}
      \centering
      \includegraphics[width=\columnwidth]{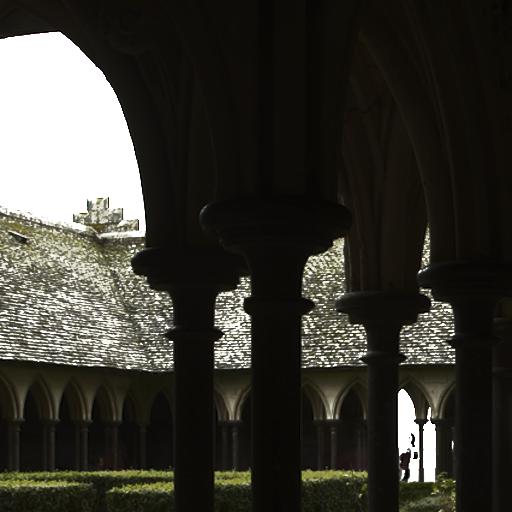}
      \caption{PMET\cite{wang2015pseudo} \label{fig:pmet5}}
    \end{subfigure}
    \begin{subfigure}[t]{0.16\hsize}
      \centering
      \includegraphics[width=\columnwidth]{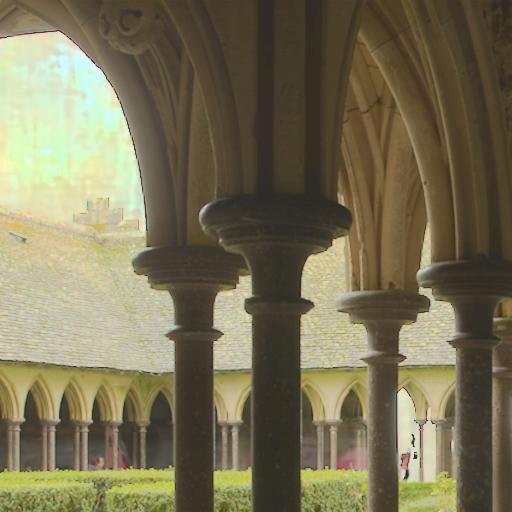}
      \caption{DrTMO\cite{endo2017deep} \label{fig:drtmo5}}
    \end{subfigure}
    \begin{subfigure}[t]{0.16\hsize}
      \centering
      \includegraphics[width=\columnwidth]{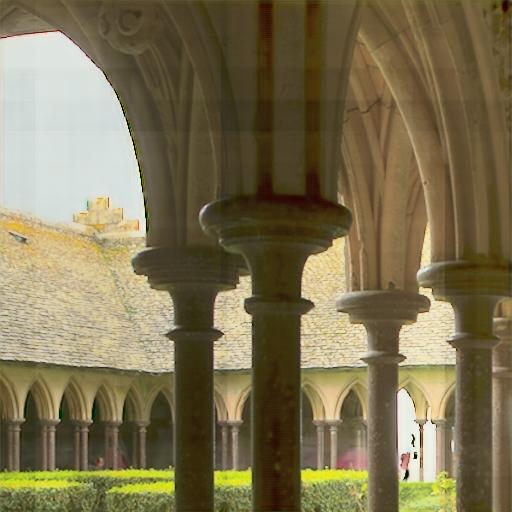}
      \caption{Proposed \label{fig:proposed5}}
    \end{subfigure}
    \begin{subfigure}[t]{0.16\hsize}
      \centering
      \includegraphics[width=\columnwidth]{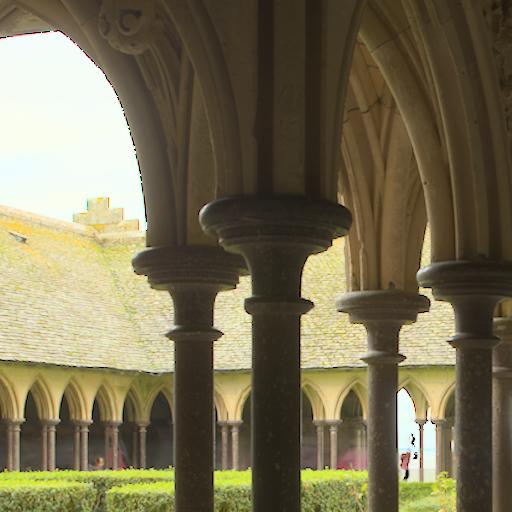}
      \caption{Ground truth $\tilde{E}$ \label{fig:original5}}
    \end{subfigure}
    \caption{Experimental Results (Image 5) \label{fig:result5}}
  \end{figure*}

  Figures \ref{fig:result1} and \ref{fig:result5} show
  examples of HDR imgaes generated by the four methods.
  Here, these images were tone-mapped from generated HDR images
  because HDR images cannot be displayed in commonly-used LDR devices.

  From Fig. \ref{fig:result1},
  it is confirmed that the proposed method produced an higher-quality HDR image
  which similar to an original HDR one $\tilde{E}$,
  than other methods.
  The image generated by DrTMO includes unnatural color distortion
  because DrTMO produced extremely bright multi-exposure images.
  Figure \ref{fig:result5} shows that
  the proposed method and DrTMO produced images which clearly represent details in dark regions.
  However, details in these regions of images generated by Direct ITMO and PMET
  are unclear.
  Therefore, the proposed method can produce higher-quality HDR images,
  which are similar to original HDR images having absolute luminance,
  than other methods.
  In addition, the proposed method has lower computational cost
  than DrTMO because DrTMO utilizes two CNNs having 3D deconvolution layers.

  Tables \ref{tab:hdrvdp} and \ref{tab:msssim} illustrate results of objective assessment
  in terms of HDR-VDP and PU encoding + MS-SSIM.
  As shown in Table \ref{tab:hdrvdp}, the proposed method provided the highest HDR-VDP scores
  in the four methods for five images.
  Since HDR-VDP evaluates HDR images in absolute-luminance scale,
  this result indicates that the proposed method can estimate absolute luminance
  with higher accuracy than the other methods.
  Moreover, the proposed method also provided the highest MS-SSIM scores
  for all images (see Table \ref{tab:msssim}).
  For these reasons, the proposed method outperforms the conventional methods
  in terms of both HDR-VDP and PU-encoding + MS-SSIM.
  \begin{table}[!t]
    \centering
    \caption{HDR-VDP-2.2 scores}
    \begin{tabular}{l|c|c|c|c}\hline\hline
      & Direct ITMO \cite{kinoshita2017fast_trans}& PMET\cite{wang2015pseudo} & DrTMO\cite{endo2017deep} & Proposed \\ \hline
      Image 1 & 32.89          & 32.89 & 47.92 & \textbf{50.94}    \\
      Image 2 & 31.19          & 31.19 & 31.41 & \textbf{35.17}    \\
      Image 3 & 28.57          & 28.57 & \textbf{45.83} & 44.86    \\
      Image 4 & \textbf{56.93} & 49.94 & 48.47 & 56.39    \\
      Image 5 & 39.38          & 39.38 & 34.33 & \textbf{48.24}    \\
      Image 6 & 33.06          & 33.06 & 48.26 & \textbf{57.61}    \\
      Image 7 & 44.77          & 44.77 & 43.82 & \textbf{61.49}    \\ \hline\hline
    \end{tabular}
    \label{tab:hdrvdp}
  \end{table}
  \begin{table}[!t]
    \centering
    \caption{PU encoding + MS-SSIM scores}
    \begin{tabular}{l|c|c|c|c}\hline\hline
      & Direct ITMO \cite{kinoshita2017fast_trans}& PMET\cite{wang2015pseudo} & DrTMO\cite{endo2017deep} & Proposed \\ \hline
      Image 1 & 0.6399 & 0.7750 & 0.4912 & \textbf{0.8828} \\
      Image 2 & 0.0599 & 0.0641 & 0.2382 & \textbf{0.8059} \\
      Image 3 & 0.7444 & 0.7393 & 0.5566 & \textbf{0.8502} \\
      Image 4 & 0.8445 & 0.5893 & 0.6624 & \textbf{0.9143} \\
      Image 5 & 0.4358 & 0.5734 & 0.1729 & \textbf{0.6641} \\
      Image 6 & 0.9156 & 0.7982 & 0.4083 & \textbf{0.9696} \\
      Image 7 & 0.5906 & 0.6997 & 0.6042 & \textbf{0.9804} \\ \hline\hline
    \end{tabular}
    \label{tab:msssim}
  \end{table}

  Those experimental results show that
  the proposed method is effective to generate high-quality HDR images
  from single LDR images.

\section{Conclusion}
  In this paper, a novel inverse tone mapping method using a CNN with
  LDR based learning was proposed.
  By using LDR images mapped by Reinhard's global operator for training,
  the proposed method enables us to estimate absolute luminance
  without specific loss functions for HDR images.
  Moreover, the proposed method has not only a higher performance
  but also a simpler network architecture, than conventional CNN based methods.
  Experimental results showed that the proposed method outperforms
  state-of-the-art inverse tone mapping methods
  in terms of visual comparison, HDR-VDP-2.2 and PU-encoding + MS-SSIM.

  In future work, we will compare computational cost of the proposed method
  with that of conventional methods.
%


\end{document}